\documentclass[runningheads]{llncs}
\usepackage{graphicx}
\usepackage{cite}
\usepackage{changepage}
\usepackage{subcaption}
\usepackage{enumitem}
\usepackage{bsymb}  
\usepackage{amsmath}
\usepackage {tikz}
\usetikzlibrary {positioning}
\usepackage[pdf]{graphviz}
\usepackage{longtable}

\newcommand\alias{\gg}
\newcommand{\eifkw}[1]{\textbf{#1}}
\newcommand{\eif}[1]{\texttt{#1}}

\newcommand{\eifcomment}[1]{{-}{-}\textsf{#1}}

\begin{document}
\title{AutoAlias: Automatic Variable-Precision Alias Analysis for Object-Oriented Programs}
%
%
\author{Victor Rivera\inst{1} \and
Bertrand Meyer\inst{1,2}}
\authorrunning{Rivera and Meyer}
%
\institute{Innopolis University, Innopolis, Russia,
\email{v.rivera@innopolis.ru}
\and
Politecnico di Milano, Milan, Italy,
\email{Bertrand.Meyer@inf.ethz.ch}}
\maketitle              
\begin{abstract}
The aliasing question (can two reference expressions point, during an execution, to the same object?) is both one of the most critical in practice, for applications ranging from compiler optimization to programmer verification, and one of the most heavily researched, with many hundreds of publications over several decades. One might then expect that good off-the-shelf solutions are widely available, ready to be plugged into a compiler or verifier. This is not the case. In practice, efficient and precise alias analysis remains an open problem.

We present a practical tool, AutoAlias, which can be used to perform automatic alias analysis for object-oriented programs. Based on the theory of ``duality semantics'', an application of Abstract Interpretation ideas, it is directed at object-oriented languages and has been implemented for Eiffel as an addition to the EiffelStudio environment. It offers variable-precision analysis, controllable through the choice of a constant that governs the number of fixpoint iterations: a higher number means better precision and higher computation time.

All the source code of AutoAlias, as well as detailed results of analyses reported in this article, are publicly available. Practical applications so far have covered a library of data structures and algorithms and a library for GUI creation. For the former, AutoAlias achieves a precision appropriate for practical purposes and execution times in the order of 25 seconds for about 8000 lines of intricate code. For the GUI library, AutoAlias produces the alias analysis in around 232 seconds for about 150000 lines of intricate code. 

\end{abstract}
\section{Introduction}
One of the most interesting questions that can be asked about a program is the aliasing question: can two given path expressions, say \eif{first\_element.next.next} and  \eif{last\_element.previous.previous.previous}, denote the same object in the run-time object structure? Alias analysis can be a key step in many applications, from compiler optimization to verification of object-oriented programs and even \cite{Meyer:14:Deadlock} deadlock analysis. 

Alias analysis has correspondingly produced an abundant literature, of which section \ref{section:RelatedWork} cites a small part. The contributions of AutoAlias, the approach described in the present work, are: a context- and flow-sensitive alias analysis technique applicable to object-oriented languages, based on a general theory of object structures; good precision, matching or exceeding the results of previous authors; an efficient implementation for object-oriented programs, currently available for Eiffel \cite{Meyer:1988}.

One of the applications of alias analysis, presented in \cite{Kogtenkov:2015, Meyer:Alias:14, BM:2010}, is change analysis, also known as ``frame inference'': what properties can an operation change? The reason alias analysis plays a key role for frame inference in object-oriented languages is that the basic property-changing operation, assignment \eif{x := e}, changes not only \eif{x} and any path expression starting with \eif{x}, such as \eif{x.a}, \eif{x.a.b} etc., but also \eif{y.x}, \eif{y.x.a} and so on for any \eif{y} that is aliased to the current object (``this''). Expanding on the original work in \cite{Kogtenkov:2015, Meyer:Alias:14, BM:2010}, we have implemented automatic frame analysis in the AutoFrame tool, based on AutoAlias. The frame analysis effort is the topic of a companion paper \cite{AutoFrame2019}.

The basic practical results are as follows, with details in section \ref{section:autoalias}. We applied AutoAlias to  a library of data structures and algorithms, EiffelBase 2, of about 8000 lines of code and 45 classes, and a significantly larger (150K-LOC) graphical (GUI) library, EiffelVision. For EiffelBase 2, to obtain a precision appropriate for practical purposes, AutoAlias takes about 25 seconds. For EiffelVision, it takes a little less than 4 minutes. In both cases the results permit detailed alias and change analysis.

The entire source code of AutoAlias is available in a public repository at \cite{AutoAlias:Impl}. The repository also contains detailed results of analyses performed in AutoAlias and reported in this article.

Some elements of this article, particularly in sections  \ref{section:math} and  \ref{alias-calc}, will at first sight appear similar to the corresponding presentations in the earlier work cited above. One of the reasons is simply to make the presentation self-contained rather than requiring the reader to go to the earlier work. More fundamentally, however, the similarity of form should not mask the fundamental differences. The mathematical model has been profoundly refined, and the implementation is completely new. The previous work is best viewed as a prototype for the present version.

Section \ref{section:RelatedWork} presents previous work. Section \ref{section:math} and  \ref{alias-calc} show the mathematical basis and theory on which AutoAlias relies. Section \ref{section:autoalias} presents the implementation of AutoAlias
its  evaluation and results. Section \ref{section:concl} concludes the work.

\section{Related Work}
\label{section:RelatedWork}
The work presented here is a continuation on the original work in \cite{Kogtenkov:2015, Meyer:Alias:14, BM:2010}. The main difference is that we present a graph-based approach to the alias analysis, whereas the previous works used a relational-based approach. The immediate advantage is in performance.

There is a considerable literature on alias analysis, in particular for compiler optimization. We only consider work that is directly comparable to the present approach. For the overall problem of alias analysis in its full generality, good surveys exist, in particular two recent ones: \cite{Smaragdakis:2015, Sridharan:2013}. AutoAlias belongs to the fairly rarefied class of approaches that are (according to the standard terminology in the field, discussed in these surveys) both:
\begin{itemize}
\item \textit{Context-sensitive}, meaning that it differentiates between executions of a given instruction in different contexts. In particular, AutoAlias is \textit{call-site-sensitive}, meaning that it does not coalesce the effects of different calls to the same routine, such as \eif{f (a1)} and \eif{f (a2)} where the routine \eif{f (a)} performs \eif{b := a}, and a context-insensitive analysis could deduce that this may alias both \eif{a1} and \eif{a2} to b and hence (wrongly) to each other.
\item \textit{Flow-sensitive}, meaning that it accounts for control flow: in \eifkw{if} \eif{c} \eifkw{then} \eif{a := x} \eifkw{else} \eif{b := x} \eifkw{end}, standard flow-insensitive analysis would report that \eif{a} and \eif{b} can get aliased to \eif{x}, but flow-sensitive analysis reports that exactly one of them will.

\end{itemize}
Both of these requirements place a much higher demand on the analysis technique.


\cite{Andersen94programanalysis} presents an efficient, inter-procedural pointer analysis for the C programming language. The analysis approximates for every variable of pointer type the set of objects it may point to during program execution. This  approach addresses C or languages of that level; the
present work has been applied to a full-fledged object-oriented language. In an O-O context some of the instructions may become unnecessary. In particular, there is no notion of plain pointers. 

A specialization of context-sensitivity is call-site sensitivity. \cite{Sharir:1981,  Shivers:1991} are the pioneers to use call sites as context. Whenever a routine gets called, the context under which the called method gets analyzed is a sequence of call sites.  Another specialization of context-sensitivity is object sensitivity \cite{Milanova:2005} and type sensitivity \cite{Smaragdakis:2011}. These approaches use object abstractions or type information as contexts. Specifically, the analysis qualifies a routine's local variables with the allocation site of the receiver object of the method call. AutoAlias follows the same spirit, however it also uses a flow-sensitive approach allowing a better precision of the analysis.  An example of flow-sensitive analysis is \cite{Hardekopf:2011}, but it too introduces imprecision, in particular in handling assignments. 

\section{The mathematical basis: object diagrams}
\label{section:math}
An object diagram is a graph where nodes represent possible objects at execution time and edges represent references variables. Let $N$ be an enumerable set of potential nodes and $T$ a set of names. An object diagram $D \langle N, T, O, R, S \rangle$ is defined by

\begin{tabular}{@{$\bullet$ }ll}
	  \eif{$O \subseteq N$} & Set of objects\\
      \eif{$R \subseteq O $}& Set of roots\\
      \eif{$S: T \tfun O \tfun \pow (O)$} & Successors (references)
\end{tabular}

Where $R$ is the set of root nodes: in O-O computations, every operation is applied to a specific object (commonly known as \textit{Current}, \textit{this} or \textit{self}). As an example, \eif{x > 10} states a property of a variable that belongs to the current object.

\begin{definition}
An alias diagram is an object diagram $G$ such that
\begin{itemize}
    \item $O$ is finite;
    \item $\mid R\mid > 1$, there exists at least one root.
\end{itemize}
\end{definition}

\noindent

\noindent
A path of edges $a,b,\ldots$ on an alias diagram is associated to the expression $a.b.\ldots$ in O-O. For a path expression $e = a.b.\ldots$, $e_G$ is the (possibly empty) set of end nodes of paths with edges $a, b, \ldots$ from any root in $G$.

An empty path is represented by \eifkw{Current} (\eifkw{Current} represents the current object in O-O computations -- also known as ``this'' or ``self''). A single-element path is written as $a$, two or more elements as $a.b.c\ldots$. We let ``.'' to also represent concatenation, e.g. if $p$ and $q$ are paths then $p.q$, $a.q$ and $p.a$ are also paths (their path concatenations). Both \eifkw{Current}\eif{.p} and \eif{p.}\eifkw{Current} mean \eif{p}.

\begin{definition}
\label{def:E}
$E$ is the set of expressions appearing in the program and its prefixes (set of all paths in $G$)
\end{definition}

\begin{definition}
\label{def:comp}
For any path $p$ in $G$, $compl (p) \subseteq E$ is the set of completion paths of $p$
$$compl (p) = \{w \in E \mid \exists_{q \in T^*} \mid w = p.q\}$$
\end{definition}

The semantics of paths is defined by value set \eif{V(p)}. The value set \eif{$V(p) \subseteq \pow (O)$} of a path \eif{p} is the set of nodes reachable from a \eif{root} through \eif{$p$}. In other words

\begin{itemize}
	\item \eif{$V(\eifkw{Current}) = R$} 
	\item \eif{$V(p.a) = S(a)(V (p))$}
\end{itemize}

\begin{definition}
\label{def:aliasing}
For any path $p$ in $G$, the set $alias_G (p) \subseteq E$ is the set of all paths that are aliased to $p$ in $G$.
$$alias_G (p) = \{q \in E \mid V(p) \binter V(q) \not = \emptyset\}$$
\end{definition}

Section \ref{sec:examples} shows some examples for a better comprehension of the definitions, as well as examples on the operations describe in the next section.

\subsection{Operations on Alias Diagrams}
\label{sec:opers}
This section describes a set of operations on an Alias Diagram $G$. Operations assume \eif{$X \subseteq O$}, \eif{$t \in T$}, and lists of the same size $l_1$, $l_2$ of expressions.  All the operations are implicitly subscripted by the name of the diagram, e.g. \eifkw{link} is really \eifkw{link$_G$}; the subscript will be omitted in the absence of ambiguity. 

\subsubsection{(Un)Linking nodes}
The set of operations shown in Table \ref{table:oper-link} is used to compute the Alias Diagram when analysing, for example, the most basic instruction for aliasing: assignment (see section \ref{alias-calc}). The effect on $G$ of an assignment is to link and unlink some of its edges.  

\begin{table}[h!]
\centering
\begin{tabular}{l|p{8cm}}
    \hline
    \textbf{operation} & \textbf{semantics} \\
    \hline \hline
    \eifkw{link}\eif{ $t:X$} & Add edge \eif{t} from every root to every member of \eif{X}. \\\hline
    \eifkw{unlink}\eif{ $t$} & Remove every edge labelled \eif{t} from roots. \\\hline
    \eifkw{unlink }\eif{$l_1$} & Shorthand for: \eifkw{unlink}\eif{ $l_1[i]$}, where $i \in 1..size(l_1)$.\\\hline
    \eifkw{relink }\eif{t:X} & Shorthand for: \eifkw{unlink }\eif{t}; \eifkw{link }\eif{t:X}. \\\hline
    \eif{$G[l_1: l_2]$} & \eifkw{link}\eif{ $l_1[i]: V_G(l_2[i])$}, where $i \in 1..size(l_1)$. \\\hline
\end{tabular}
\caption{Operations for linking.}
\label{table:oper-link}
\end{table}
The last operation is particularly useful to link actual arguments to formal arguments in a feature call (for more details, see rule \texttt{AC-UQCall} in section \ref{alias-calc}).

\subsubsection{Rooting nodes}
The operation shown in Table \ref{table:oper-root} is used to compute the effect on $G$ when analysing qualified feature calls (for more details, see section \ref{alias-calc}, rule \texttt{AC\_QCall}). In O-O computations, calls are applied to a specific object. For instance, a call to the feature \eif{set\_x (3)} is applied to the current object (also know as \textit{this} or \textit{self}) and its effect might change its state. A call to the qualified feature \eif{y.set\_x (3)} might change the state of the object $y$. When a qualify call occurs, it is necessary to change the root of $G$ so the effect of executing the feature modifies the corresponding object. 

\begin{table}[h!]
\centering
\begin{tabular}{l|p{8cm}}
    \hline
    \textbf{operation} & \textbf{semantics} \\
    \hline \hline
   \eifkw{reroot }\eif{X} & Replace \eif{$R$} by \eif{$X$}. \\\hline
   
\end{tabular}
\caption{Operation for rooting.}
\label{table:oper-root}
\end{table}

This operation (along with dot distribution -- see section \ref{sec:dot-dist}) allows the analysis to be call-site sensitive.

\subsubsection{Including nodes}
The operation shown in table \ref{table:oper-incl} is used to compute the effect on $G$ when analysing a creation instruction in the code (for more details, see section \ref{alias-calc}, rule \texttt{AC\_New}). The operations add a new object that does not currently exist in $G$.

\begin{table}[h!]
\centering
\begin{tabular}{l|p{8cm}}
    \hline
    \textbf{operation} & \textbf{semantics} \\
    \hline \hline
    \eifkw{include }\eif{o} & Choose \eif{$o \in N$} where \eif{$o \not \in O$} and add it to \eif{$O$}.\\\hline
    
\end{tabular}
\caption{Operations for inclusion.}
\label{table:oper-incl}
\end{table}

\subsubsection{New Alias Diagrams}
The operations shown in table \ref{table:oper-union} are used to compute the effect on $G$ when analysing a conditional or loop instruction in the code. 

\begin{table}[h!]
\centering
\begin{tabular}{l|p{8cm}}
    \hline
    \textbf{operation} & \textbf{semantics} \\
    \hline \hline
    $G_1 \bunion G_2 $ & Yields the union of sets defined by $G_1$ and $G_2$. \\
    &$G = G_1 \bunion G_2$, where $N = N_1 \bunion N_2$, $R = R_1 \bunion R_2$ and $S = S_1 \bunion S_2$\\\hline
    
    $clone (G) $ & Yields an Alias Diagram $G'$ defined over a new set of objects $O'
    \subseteq N - O$ and isomorphic to $G$.\\\hline
    
\end{tabular}
\caption{Alias Diagrams union.}
\label{table:oper-union}
\end{table}

Rules \texttt{AC\_Cond} and \texttt{AC\_Loop} use these operations to create new Alias Diagrams for each branch (either in conditionals or loops). This allows the analysis to be flow-sensitive when analysing conditionals and loops (see section \ref{alias-calc} for more details).

\subsection{Generalization of dot distribution over Alias Diagrams}
\label{sec:dot-dist}
The basic mechanism of Object Oriented computations is feature call. All computations are achieved by calling certain features on a certain object. Consider \eif{x.f}, this particular call means \textit{apply feature \eif{f} to the object attached to \eif{x}}. Alias diagrams are built upon this mechanism. Authors in \cite{Meyer2014} introduce the notion of ``distributed dot'' that distributes the period of O-O programming over a list, a set or a relation; for example, \eif{$x\bullet [u,v,w]$} denotes the list \eif{[x.u,x.v,x.w]}. We extend the mechanism to dot distribution over Alias Diagrams.

\begin{definition}
    For an alias diagram \eif{G}, \eif{$x\bullet G$} adds a back-pointer \eif{$x'$} from \eif{$x$} to each element in $R$, the roots of $G$. In other words, the effect of \eif{$x\bullet G$} is  \eif{$S \bunion \{(x', o, R) \mid o \in V(x)\}$}.
\end{definition}

\begin{figure}[h!]
\centering
\begin {tikzpicture}[-latex ,auto ,node distance =2 cm  ,on grid ,
semithick , state/.style ={ circle ,top color =white ,  draw}]

        \node[state] (n0)                    {\underline{$n_0$}};
        \node[state] (n1) [right =of n0] {$n_1$};
        \node[state] (n2) [below =of n1] {$n_2$};
        \path  
            (n0) edge [] node {$a$} (n1)
            (n0) edge [] node {$b$} (n2);
        \path 
            (n1) edge [loop above] node {$d$} (n1);
            
        \node[state] (n20)  [right =of n1, xshift=1.5cm]  {\underline{$n_0$}};
        \node[state] (n21) [right =of n20] {$n_1$};
        \node[state] (n22) [below =of n21] {$n_2$};
        \path 
            (n20) edge [] node {$a$} (n21)
            (n20) edge [] node {$b$} (n22)
            ;
        \path 
            (n22) edge [bend left, red, below] node {$b'$} (n20)
            ;
        \path 
            (n21) edge [loop above] node {$d$} (n21);
    \end{tikzpicture}
\caption{On the left, Alias Diagram $G$. On the right the Alias Diagram \eif{$b\bullet G$}.}
\label{fig:dot_distribution}
\end{figure}
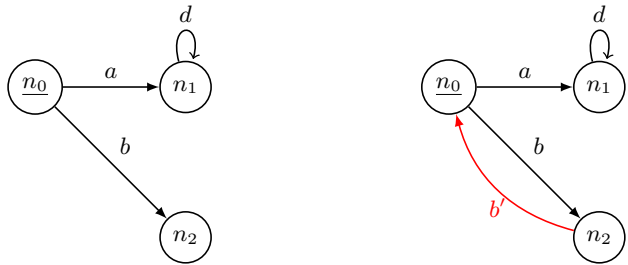

\noindent
Figure \ref{fig:dot_distribution} depicts the effect of performing dot distribution over a graph.

This rule enables the analysis of alias diagrams to transpose the context of a call to the context of the caller since it may depend and act on values and properties that are set by the object that launched the current call. 

\subsection{Examples}
\label{sec:examples}
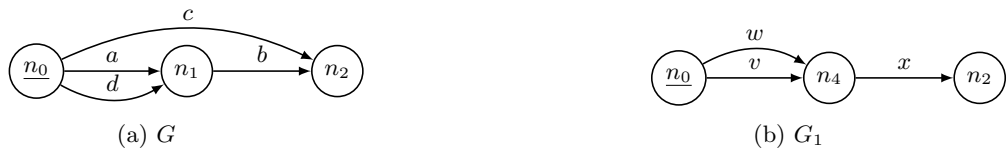
\begin{figure}[h!]
\centering
\begin{subfigure}[b]{0.3\textwidth}
    \begin {tikzpicture}[-latex ,auto ,node distance =2 cm  ,on grid ,
semithick , state/.style ={ circle ,top color =white ,  draw, minimum width =0.6 cm}]

    \node[state] (A)                    {\underline{$n_0$}};
\node[state] (B) [right =of A] {$n_1$};
\node[state] (C) [right =of B] {$n_2$};
\path (A) edge [] node {$a$} (B);
\path (A) edge [bend right] node {$d$} (B);
\path (B) edge [] node {$b$} (C);
\path (A) edge [bend left =25] node {$c$} (C);
    
\end{tikzpicture}
\caption{$G$}
\label{fig:alias_graph:g}
\end{subfigure}
    \hfill 
\begin{subfigure}[b]{0.3\textwidth}
    \begin {tikzpicture}[-latex ,auto ,node distance =2 cm  ,on grid ,
semithick , state/.style ={ circle ,top color =white ,  draw, minimum width =0.6 cm}]

    \node[state] (n20)  [right =of C, xshift=1.5cm]  {\underline{$n_0$}};
        \node[state] (n4) [right =of n20] {$n_4$};
        \node[state] (n22) [right =of n4] {$n_2$};
        \path 
            (n20) edge [] node {$v$} (n4)
            (n20) edge [bend left] node {$w$} (n4)
            ;
        \path 
            (n4) edge [] node {$x$} (n22);

\end{tikzpicture}
\caption{$G_1$}
\label{fig:alias_graph:g1}
\end{subfigure}
\caption{Examples of Alias diagrams.}
\label{fig:alias_graph}
\end{figure}

Figure \ref{fig:alias_graph} is the graphical representation of possible Alias Diagrams. The set of nodes for $G$ (see figure \ref{fig:alias_graph:g1}) is $N_G = \{\underline{n_0}, n_1, n_2\}$ and the successors $T_G = \{(a, \underline{n_0}, \{n_1\}), (d, \underline{n_0}, \{n1\}), (c, \underline{n_0}, \{n_2\}), (b, n_1, \{n_2\})\}$. Alias Diagrams have at least one root, $R_G = \{n_0\}$ (we use underlined nodes to graphically represent the set of roots). 

The set of all expressions in the graph is $E_G = \{a, b, c, d, a.b, d.b\}$. This set is used to get the set of expression completion. This set is not particularly interesting for aliasing, however it is an important definition to be used in the Framing problem: the problem of inferring all program locations that might change. As an example, the completion path of $d$ is $compl_G (d) = \{d.b\}$ and the completion path of $a.b$ is $compl_G (a.b) = \emptyset$. Set $E$ is also used to compute aliasing, see definition \ref{def:aliasing}. As an example, the expressions aliased to $c$ are $alias_G (c) = \{c, a.b, d.b\}$.

Table \ref{table:examples} shows the application of the different operations defined is section \ref{sec:opers} on the Alias Diagrams $G$ and $G_1$ depicted in figure \ref{fig:alias_graph}. 

\begin{longtable}{p{5cm}||p{5cm}}
  \textbf{Operation} & \textbf{Effect on $G$ or $G_1$ (in figure \ref{fig:alias_graph})}  \\ \hline \hline
  \eifkw{link$_G$}\eif{ $f:\{n_1\}$} &
    
   \begin {tikzpicture}[-latex ,auto ,node distance =2 cm  ,on grid ,
semithick , state/.style ={ circle ,top color =white ,  draw, minimum width =0.6 cm}]

\node[state] (A)                    {\underline{$n_0$}};
\node[state] (B) [right =of A] {$n_1$};
\node[state] (C) [right =of B] {$n_2$};
\path (A) edge [] node {$a$} (B);
\path (A) edge [bend right] node {$d$} (B);
\path (B) edge [] node {$b$} (C);
\path (A) edge [bend left =25] node {$c$} (C);
\path (A) edge [bend right =100] node {$f$} (B);
\end{tikzpicture}
 \\ \hline \hline
  \eifkw{unlink$_G$}\eif{ $d$} &
    
   \begin {tikzpicture}[-latex ,auto ,node distance =2 cm  ,on grid ,
semithick , state/.style ={ circle ,top color =white ,  draw, minimum width =0.6 cm}]

\node[state] (A)                    {\underline{$n_0$}};
\node[state] (B) [right =of A] {$n_1$};
\node[state] (C) [right =of B] {$n_2$};
\path (A) edge [] node {$a$} (B);
\path (B) edge [] node {$b$} (C);
\path (A) edge [bend left =25] node {$c$} (C);
\end{tikzpicture}

 \\ \hline \hline
  \eifkw{unlink$_G$}\eif{ $\{c, d\}$} &
    
   \begin {tikzpicture}[-latex ,auto ,node distance =2 cm  ,on grid ,
semithick , state/.style ={ circle ,top color =white ,  draw, minimum width =0.6 cm}]

\node[state] (A)                    {\underline{$n_0$}};
\node[state] (B) [right =of A] {$n_1$};
\node[state] (C) [right =of B] {$n_2$};
\path (A) edge [] node {$a$} (B);
\path (B) edge [] node {$b$} (C);
\end{tikzpicture}

 \\ \hline \hline
  \eifkw{relink$_G$}\eif{ $a:\{n_2\}$} &
    
   \begin {tikzpicture}[-latex ,auto ,node distance =2 cm  ,on grid ,
semithick , state/.style ={ circle ,top color =white ,  draw, minimum width =0.6 cm}]

\node[state] (A)                    {\underline{$n_0$}};
\node[state] (B) [right =of A] {$n_1$};
\node[state] (C) [right =of B] {$n_2$};
\path (A) edge [bend right =25, below] node {$a$} (C);
\path (A) edge [] node {$d$} (B);
\path (B) edge [] node {$b$} (C);
\path (A) edge [bend left =25] node {$c$} (C);
\end{tikzpicture}

 \\ \hline \hline
  \eifkw{G[$[c,d] : [a,b]$]}\par where $V_G (a) = \{n_1\}$ and $V_G (b) = \{n_2\}$&
  
   \begin {tikzpicture}[-latex ,auto ,node distance =2 cm  ,on grid ,
semithick , state/.style ={ circle ,top color =white ,  draw, minimum width =0.6 cm}]

\node[state] (A)                    {\underline{$n_0$}};
\node[state] (B) [right =of A] {$n_1$};
\node[state] (C) [right =of B] {$n_2$};
\path (A) edge [] node {$a$} (B);
\path (A) edge [bend right] node {$d$} (B);
\path (B) edge [] node {$b$} (C);
\path (A) edge [bend left =25] node {$c$} (C);
\path (A) edge [bend right =50, below] node {$c$} (B);
\path (A) edge [bend right =50, below] node {$d$} (C);
\end{tikzpicture}

 \\ \hline \hline
  \eifkw{reroot$_G$}\eif{ $\{n_2\}$} \par Graphically, roots are shown as underlined nodes.  &
    
   \begin {tikzpicture}[-latex ,auto ,node distance =2 cm  ,on grid ,
semithick , state/.style ={ circle ,top color =white ,  draw, minimum width =0.6 cm}]

\node[state] (A)                    {$n_0$};
\node[state] (B) [right =of A] {$n_1$};
\node[state] (C) [right =of B] {\underline{$n_2$}};
\path (A) edge [] node {$a$} (B);
\path (A) edge [bend right] node {$d$} (B);
\path (B) edge [] node {$b$} (C);
\path (A) edge [bend left =25] node {$c$} (C);
\end{tikzpicture}

 \\ \hline \hline
  \eifkw{include$_G$}\eif{ $f:\{n_n\}$} &
    
   \begin {tikzpicture}[-latex ,auto ,node distance =2 cm  ,on grid ,
semithick , state/.style ={ circle ,top color =white ,  draw, minimum width =0.6 cm}]

\node[state] (A)     {\underline{$n_0$}};
\node[state] (B) [right =of A] {$n_1$};
\node[state] (C) [right =of B] {$n_2$};
\node[state] (D) [below =1cm of A] {$n_n$};
\path (A) edge [] node {$a$} (B);
\path (A) edge [bend right] node {$d$} (B);
\path (B) edge [] node {$b$} (C);
\path (A) edge [bend left =25] node {$c$} (C);
\end{tikzpicture}

 \\ \hline \hline
  $G \bunion G_1$ &
    
   \begin {tikzpicture}[-latex ,auto ,node distance =2 cm  ,on grid ,
semithick , state/.style ={ circle ,top color =white ,  draw, minimum width =0.6 cm}]

\node[state] (A)                    {\underline{$n_0$}};
\node[state] (B) [right =of A] {$n_1$};
\node[state] (C) [right =of B] {$n_2$};
\node[state] (D) [below =1cm of B] {$n_4$};
\path (A) edge [] node {$a,d$} (B);
\path (A) edge [below] node {$v,w$} (D);
\path (A) edge [bend left =25] node {$c$} (C);
\path (B) edge [] node {$b$} (C);
\path (D) edge [below] node {$x$} (C);
\end{tikzpicture}

 \\ \hline \hline
  \eifkw{clone ($G$)} &
    
   \begin {tikzpicture}[-latex ,auto ,node distance =2 cm  ,on grid ,
semithick , state/.style ={ circle ,top color =white ,  draw, minimum width =0.6 cm}]

\node[state] (A)                    {\underline{$n_0'$}};
\node[state] (B) [right =of A] {$n_1'$};
\node[state] (C) [right =of B] {$n_2'$};
\path (A) edge [] node {$a$} (B);
\path (A) edge [bend right] node {$d$} (B);
\path (B) edge [] node {$b$} (C);
\path (A) edge [bend left =25] node {$c$} (C);
\end{tikzpicture}

    \\ \hline
    
\caption{Operations on Alias Diagrams}
\label{table:examples}
\end{longtable}

\section{The Alias Calculus}
\label{alias-calc}
The alias calculus, is a set of rules defining the effect of executing an instruction on the aliasing that may exist between expressions. Each of these rules gives, for an instruction \eif{$p$} of a given kind and an alias diagram
\eif{$G$} that holds in the initial state, the value of \eif{$G \alias p$}, the alias diagram that holds after the execution of \eif{$p$}.

\subsection{The programming language}

The programming language figuring in the rules of the Alias Calculus given below is a common-core subset of modern object-oriented languages, including the fundamental constructs found, with varying syntax and other details, in Java, C\#, Eiffel, C++ and others: respectively reference assignment, composition (sequencing), object creation (\textbf{new}), conditional, loop, unqualified call and qualified call. As a result, the present work applies to any O-O language, with possible fine-tuning to account for individual differences; and so potentially does AutoAlias, although so far we have applied it to Eiffel only.

Following the earlier work, the programming language does not have a real conditional instruction \eifkw{if} \eif{c} \eifkw{then} \eif{p} \eifkw{else} \eif{q} \eifkw{end}, but only a non-deterministic choice written \eifkw{then} \eif{p} \eifkw{else} \eif{q} \eifkw{end}, which executes either \eif{p} or \eif{q}. The loop construct similarly does not list a condition: \eifkw{loop} \eif{p} \eifkw{end} executes $b$ any number of times including zero. Ignoring conditions causes a potential loss of precision; as a trivial example, ignoring the condition in \eifkw{if} \eif{n $>$ n + 1} \eifkw{then} \eif{a := b} \eifkw{else} \eif{a := c} \eifkw{end} leads to concluding wrongly (that is to say, soundly but with a loss of precision) that \eif{a} may become aliased to \eif{b}. The Alias Calculus only knows about the object diagram and its reference structure; other properties, such as arithmetic properties in this example, are beyond its reach. Unlike the previous version of this work, however, AutoAlias can now deal with a limited set of properties which indeed pertain to the object structure. For that reason the language now includes a construct

\eifkw{if} \eif{c: p}

with the semantics of doing nothing (``skip'') if it is known for sure that \eif{c} does not hold, and otherwise (that is to say, if the analysis can deduce that \eif{c} holds, or cannot draw a conclusion) to execute \eif{p}. So the standard conditional instruction of programming languages can be handled in the Alias Calculus as \eifkw{then} (\eifkw{if} \eif{c: p}) \eifkw{else} (\eifkw{if} $\neg$ \eif{c: q}) \textbf{end}. At present AutoAlias has semantics for simple conditions on references such as \eif{e = f} (equality) and \eif{e /= f} (inequality) for path expressions ($a.b.c\ldots$) $e$ and $f$.

\subsection{The calculus}
Rules of the calculus are shown in Table \ref{table:alias-calculus}. The table shows the name of the rule, the rule $G \alias p$, where $G$ is an Alias Diagram and $p$ is an instruction, and its semantics, the effect of executing $p$ on the aliasing that may exist between expressions.

\begin{table}[h!]
\centering
\begin{tabular}{l|l|l}
    \hline
    Rule Name &Rule & Semantics \\ \hline \hline
	\texttt{AC-Assg} & \eif{$G \alias $} (\eif{t := s}) & = \eifkw{relink }\eif{$t:V_G(s)$}\\\hline
    
    \texttt{AC-Comp} &
    \eif{$G \alias $} (\eif{p;q})&= (\eif{$G \alias p$}) \eif{$ \alias q$}\\ \hline
    
    \texttt{AC-New} & \eif{$G \alias $} (\eifkw{create }\eif{ t}) &= \eifkw{include }\eif{n; }\eifkw{link}\eif{ $t:\{n\}$}\\ \hline
    
    \texttt{AC-Cond} & \eif{$G \alias$} (\eifkw{then} \eif{ p } \eifkw{else} \eif{ q } \eifkw{end}) &=
			(\eif{$G \alias p$}) $\cup$ (\eifkw{clone}\eif{ ($G \alias q$}))\\\hline
    
    \texttt{AC-Loop} & \eif{$G \alias $ }(\eifkw{loop} \eif{ p } \eifkw{end})&= $\bigcup\limits_{i \in \mathbb{N}}$ (\eif{$G_i \alias p$})\\\hline

    \texttt{AC-UQCall} &\eif{$G \alias $} (\eifkw{call} \eif{ f (l)})&= (\eif{$G[f^\bullet : l] \alias {\mid}f{\mid}$})\\ \hline

    \texttt{AC-QCall} & \eif{$G \alias $} (\eif{x.}\eifkw{call} \eif{ f (l)})&
    $\eifkw{reroot}_{
        ((
        	\eifkw{reroot}_{x'\bullet G} V(x)
            \alias 
            \eifkw{call}  f(x'.l)
        ))} R_G
        $
    
    \\\hline
	\end{tabular}
\caption{The Alias Calculus. A set of rules over an Alias Diagram $G$ and their semantics.}
\label{table:alias-calculus}
\end{table}

\subsubsection{Assignments. }
Rule \texttt{AC-Assg} deals with assignments: the main instruction that creates aliasing. Figure \ref{ex:assg} shows an example on how the rule is applied to the instruction \eif{a := b} on the Alias Diagram $G$ in figure \ref{ex:assg:a}. The semantics for \eif{$G \alias $} (\eif{a := b}) is \eifkw{relink }\eif{$a:V_G(b)$} which is shorthand for applying \eifkw{unlink }$a$ (figure \ref{ex:assg:b}) then \eifkw{link }\eif{$a:V_G(b)$} (figure \ref{ex:assg:c}).

\begin{figure}[h!]
\centering
\begin{subfigure}[b]{0.3\textwidth}
    \begin {tikzpicture}[-latex ,auto ,node distance =2 cm  ,on grid ,
semithick , state/.style ={ circle ,top color =white ,  draw, minimum width =0.6 cm}]

    \node[state] (A)                    {\underline{$n_0$}};
    \node[state] (B) [right =of A] {$n_1$};
    \node[state] (C) [below =1.5cm of A] {$n_2$};
    \node[state] (D) [right =of C] {$n_3$};
    \path (A) edge [] node {$a$} (B);
    \path (A) edge [] node {$b$} (C);
    \path (A) edge [below] node {$x$} (D);
    
\end{tikzpicture}
\caption{}
\label{ex:assg:a}
\end{subfigure}
    \hfill 
\begin{subfigure}[b]{0.3\textwidth}
    \begin {tikzpicture}[-latex ,auto ,node distance =2 cm  ,on grid ,
semithick , state/.style ={ circle ,top color =white ,  draw, minimum width =0.6 cm}]

    \node[state] (A)                    {\underline{$n_0$}};

    \node[state] (C) [below =1.5cm of A] {$n_2$};
    \node[state] (D) [right =of C] {$n_3$};

    \path (A) edge [] node {$b$} (C);
    \path (A) edge [below] node {$x$} (D);

\end{tikzpicture}
\caption{}
\label{ex:assg:b}
\end{subfigure}
    \hfill 
\begin{subfigure}[b]{0.3\textwidth}
    \begin {tikzpicture}[-latex ,auto ,node distance =2 cm  ,on grid ,
semithick , state/.style ={ circle ,top color =white ,  draw, minimum width =0.6 cm}]

    \node[state] (A)                    {\underline{$n_0$}};
    \node[state] (C) [below =1.5cm of A] {$n_2$};
    \node[state] (D) [right =of C] {$n_3$};
    \path (A) edge [bend left] node {$a$} (D);
    \path (A) edge [] node {$b$} (C);
    \path (A) edge [below] node {$x$} (D);

\end{tikzpicture}
\caption{}
\label{ex:assg:c}
\end{subfigure}
\caption{Applying rule \texttt{AC-Assg}}
\label{ex:assg}
\end{figure}

\subsubsection{Composition. } Rule \texttt{AC-Comp} deals with a compound of instructions (e.g. the set of instructions in a routine). Figure \ref{ex:comp} shows an example on how the rule is applied to the instruction \eif{a := x;b := x} on the Alias Diagram $G$ in figure \ref{ex:comp:a}. The semantics for \eif{$G \alias $} (\eif{a := x; b := x}) is 
 $$\overbrace{\underbrace{(G \alias  (\eif{a := x))}}_\text{see figure \ref{ex:comp:b}}
 \alias  (b := x)}^\text{see figure \ref{ex:comp:c}}$$. 

\begin{figure}[h!]
\centering
\begin{subfigure}[b]{0.3\textwidth}
    \begin {tikzpicture}[-latex ,auto ,node distance =2 cm  ,on grid ,
semithick , state/.style ={ circle ,top color =white ,  draw, minimum width =0.6 cm}]

    \node[state] (A)                    {\underline{$n_0$}};
    \node[state] (B) [right =of A] {$n_1$};
    \node[state] (C) [below =1.5cm of A] {$n_2$};
    \node[state] (D) [right =of C] {$n_3$};
    \path (A) edge [] node {$a$} (B);
    \path (A) edge [] node {$b$} (C);
    \path (A) edge [below] node {$x$} (D);
    
\end{tikzpicture}
\caption{}
\label{ex:comp:a}
\end{subfigure}
    \hfill 
\begin{subfigure}[b]{0.3\textwidth}
    \begin {tikzpicture}[-latex ,auto ,node distance =2 cm  ,on grid ,
semithick , state/.style ={ circle ,top color =white ,  draw, minimum width =0.6 cm}]

    \node[state] (A)                    {\underline{$n_0$}};

    \node[state] (C) [below =1.5cm of A] {$n_2$};
    \node[state] (D) [right =of C] {$n_3$};

    \path (A) edge [] node {$b$} (C);
    \path (A) edge [bend left] node {$a$} (D);
    \path (A) edge [below] node {$x$} (D);

\end{tikzpicture}
\caption{}
\label{ex:comp:b}
\end{subfigure}
    \hfill 
\begin{subfigure}[b]{0.3\textwidth}
    \begin {tikzpicture}[-latex ,auto ,node distance =2 cm  ,on grid ,
semithick , state/.style ={ circle ,top color =white ,  draw, minimum width =0.6 cm}]

    \node[state] (A)                    {\underline{$n_0$}};
    
    \node[state] (D) [right =of C] {$n_3$};
    \path (A) edge [bend left] node {$a$} (D);
     \path (A) edge [bend right, below] node {$b$} (D);
    
    \path (A) edge [below] node {$x$} (D);

\end{tikzpicture}
\caption{}
\label{ex:comp:c}
\end{subfigure}
\caption{Applying rule \texttt{AC-Comp}}
\label{ex:comp}
\end{figure}

\subsubsection{Creation. } Rule \texttt{AC-New} deals with  the creation of new objects. Figure \ref{ex:new} shows an example on how the rule is applied to the instruction \eifkw{create }\eif{x} (also known as \eif{x = }\eifkw{new }\eif{T();} in some Programming Languages) on the Alias Diagram $G$ in figure \ref{ex:new:a}. The semantics for \eif{$G \alias $} (\eifkw{create}\eif{ x}) is \eifkw{include }\eif{$n_4$} (figure \ref{ex:new:b} -- $n_4$ is just a new node on $G$) then \eifkw{relink }\eif{$x:\{n_4\}$} (figure \ref{ex:new:c}).

\begin{figure}[h!]
\centering
\begin{subfigure}[b]{0.3\textwidth}
    \begin {tikzpicture}[-latex ,auto ,node distance =2 cm  ,on grid ,
semithick , state/.style ={ circle ,top color =white ,  draw, minimum width =0.6 cm}]

    \node[state] (A)                    {\underline{$n_0$}};
    \node[state] (B) [right =of A] {$n_1$};
    \node[state] (C) [below =1.5cm of A] {$n_2$};
    \node[state] (D) [right =of C] {$n_3$};
    \path (A) edge [] node {$a$} (B);
    \path (A) edge [] node {$b$} (C);
    \path (A) edge [] node {$x$} (D);
    
\end{tikzpicture}
\caption{}
\label{ex:new:a}
\end{subfigure}
    \hfill 
\begin{subfigure}[b]{0.3\textwidth}
    \begin {tikzpicture}[-latex ,auto ,node distance =2 cm  ,on grid ,
semithick , state/.style ={ circle ,top color =white ,  draw, minimum width =0.6 cm}]

    \node[state] (A)                    {\underline{$n_0$}};
    \node[state] (B) [right =of A] {$n_1$};
    \node[state] (C) [below =1.5cm of A] {$n_2$};
    \node[state] (D) [right =of C] {$n_3$};
    \node[state] (E) [right =1cm of C] {$n_4$};
    \path (A) edge [] node {$a$} (B);
    \path (A) edge [] node {$b$} (C);
    \path (A) edge [] node {$x$} (D);

\end{tikzpicture}
\caption{}
\label{ex:new:b}
\end{subfigure}
    \hfill 
\begin{subfigure}[b]{0.3\textwidth}
    \begin {tikzpicture}[-latex ,auto ,node distance =2 cm  ,on grid ,
semithick , state/.style ={ circle ,top color =white ,  draw, minimum width =0.6 cm}]

    \node[state] (A)                    {\underline{$n_0$}};
    \node[state] (B) [right =of A] {$n_1$};
    \node[state] (C) [below =1.5cm of A] {$n_2$};
    \node[state] (D) [right =of C] {$n_4$};
    \path (A) edge [] node {$a$} (B);
    \path (A) edge [] node {$b$} (C);
    \path (A) edge [] node {$x$} (D);

\end{tikzpicture}
\caption{}
\label{ex:new:c}
\end{subfigure}
\caption{Applying rule \texttt{AC-New}}
\label{ex:new}
\end{figure}

\subsubsection{Conditionals. } Rule \texttt{AC-Cond} deals with conditionals. The rule does not take into consideration the condition, rather treats the instruction as a non-deterministic choice. The rule assumes the command-query separation principle \cite{Meyer:1997:OSC}: asking a question should not change the answer. In other words, the rule assumes that functions being called in the condition are pure. Figure \ref{ex:cond} shows an example on how the rule is applied to the instruction \eifkw{then}\eif{ a := x}\eifkw{ else}\eif{ b := x}\eifkw{ end} on the Alias Diagram $G$ in figure \ref{ex:cond:a}. The semantics for \eif{$G \alias $} (\eifkw{then}\eif{ a := x}\eifkw{ else}\eif{ b := x}\eifkw{ end}) is 

$$\overbrace{\underbrace{G \alias (a := x)}_\text{see figure \ref{ex:cond:b}} \bunion \underbrace{\eifkw{clone} (\underbrace{G \alias (b := x)}_\text{see figure \ref{ex:cond:c}})}_\text{see figure \ref{ex:cond:d}}}^\text{see figure \ref{ex:cond:e}}$$

\begin{figure}[h!]
\centering
\begin{subfigure}[b]{0.3\textwidth}
    \begin {tikzpicture}[-latex ,auto ,node distance =2 cm  ,on grid ,
semithick , state/.style ={ circle ,top color =white ,  draw, minimum width =0.6 cm}]

    \node[state] (A)                    {\underline{$n_0$}};
    \node[state] (B) [right =of A] {$n_1$};
    \node[state] (C) [below =1.5cm of A] {$n_2$};
    \node[state] (D) [right =of C] {$n_3$};
    \path (A) edge [] node {$a$} (B);
    \path (A) edge [] node {$b$} (C);
    \path (A) edge [below] node {$x$} (D);
    
\end{tikzpicture}
\caption{}
\label{ex:cond:a}
\end{subfigure}
    \hfill 
\begin{subfigure}[b]{0.3\textwidth}
    \begin {tikzpicture}[-latex ,auto ,node distance =2 cm  ,on grid ,
semithick , state/.style ={ circle ,top color =white ,  draw, minimum width =0.6 cm}]

    \node[state] (A)                    {\underline{$n_0$}};
    \node[state] (C) [below =1.5cm of A] {$n_2$};
    \node[state] (D) [right =of C] {$n_3$};
    \path (A) edge [bend left] node {$a$} (D);
    \path (A) edge [] node {$b$} (C);
    \path (A) edge [below] node {$x$} (D);

\end{tikzpicture}
\caption{}
\label{ex:cond:b}
\end{subfigure}
    \hfill 
\begin{subfigure}[b]{0.3\textwidth}
    \begin {tikzpicture}[-latex ,auto ,node distance =2 cm  ,on grid ,
semithick , state/.style ={ circle ,top color =white ,  draw, minimum width =0.6 cm}]

    \node[state] (A)                    {\underline{$n_0$}};
    \node[state] (B) [right =of A] {$n_1$};
    \node[state] (D) [right =of C] {$n_3$};
    \path (A) edge [] node {$a$} (B);
    \path (A) edge [bend right, below] node {$b$} (D);
    \path (A) edge [below] node {$x$} (D);

\end{tikzpicture}
\caption{}
\label{ex:cond:c}
\end{subfigure}
    \hfill 
\begin{subfigure}[b]{0.3\textwidth}
    \begin {tikzpicture}[-latex ,auto ,node distance =2 cm  ,on grid ,
semithick , state/.style ={ circle ,top color =white ,  draw, minimum width =0.6 cm}]

    \node[state] (A)                    {\underline{$n_0'$}};
    \node[state] (B) [right =of A] {$n_1'$};
    \node[state] (D) [right =of C] {$n_3'$};
    \path (A) edge [] node {$a$} (B);
    \path (A) edge [bend right, below] node {$b$} (D);
    \path (A) edge [below] node {$x$} (D);

\end{tikzpicture}
\caption{clone of figure \ref{ex:cond:c}}
\label{ex:cond:d}
\end{subfigure}
    \quad 
\begin{subfigure}[b]{0.3\textwidth}
    \begin {tikzpicture}[-latex ,auto ,node distance =2 cm  ,on grid ,
semithick , state/.style ={ circle ,top color =white ,  draw, minimum width =0.6 cm}]

    \node[state] (A)                    {\underline{$n_0$}};
    \node[state] (C) [below =1.5cm of A] {$n_2$};
    \node[state] (D) [right =of C] {$n_3$};
    \path (A) edge [bend left] node {$a$} (D);
    \path (A) edge [] node {$b$} (C);
    \path (A) edge [below] node {$x$} (D);
    
    \node[state] (A2)  [left =4cm of A]{\underline{$n_0'$}};
    \node[state] (B2) [right =of A2] {$n_1'$};
    \node[state] (D2) [below =1.5cm of B2] {$n_3'$};
    \path (A2) edge [] node {$a$} (B2);
    \path (A2) edge [bend right, below] node {$b$} (D2);
    \path (A2) edge [below] node {$x$} (D2);

\end{tikzpicture}
\caption{union of Alias Diagram in figures \ref{ex:cond:b} and \ref{ex:cond:d}}
\label{ex:cond:e}
\end{subfigure}
\caption{Applying rule \texttt{AC-Cond}}
\label{ex:cond}
\end{figure}

The semantics of rule \texttt{AC-Cond} is sound but adds imprecision. This negatively affects the performance of the computation. Some improvements of the rule can be introduced: each branch of a conditional will only change a small part of the diagram, thus there is not need to clone the common parts; the \eifkw{clone} operation can be changed to clone only the source node. Figure \ref{ex:cond:f} depicts the result of the union of the alias diagram in figure \ref{ex:cond:b} and figure \ref{ex:cond:d} applying the optimization in operation \eifkw{clone}.

\begin{figure}[h!]
\centering
\begin {tikzpicture}[-latex ,auto ,node distance =2 cm  ,on grid ,
semithick , state/.style ={ circle ,top color =white ,  draw}]

        \node[state] (n0)                    {\underline{$n_0$}};
        \node[state] (n2) [below =of n0] {$n_2$};
        \node[state] (n3) [right =of n2] {$n_3$};
        \node[state] (n1) [right =of n3] {$n_1$};
        \node[state] (n0p) [above =of n1] {\underline{$n_0'$}};

        \path  
            (n0) edge [left] node {$b$} (n2)
            (n0) edge [bend left] node {$a$} (n3)
            (n0) edge [bend right] node {$x$} (n3);
        \path 
            (n0p) edge [] node {$a$} (n1)
            (n0p) edge [bend left, above] node {$b$} (n3)
            (n0p) edge [bend right, above] node {$x$} (n3);
    \end{tikzpicture}
\caption{\eif{$G \alias $} (\eifkw{then}\eif{ a := x}\eifkw{ else}\eif{ b := x}\eifkw{ end}) optimized.}
\label{ex:cond:f}
\end{figure}

The \texttt{AC-Cond} rule (and its optimization) is sound. The example shown in figure \ref{ex:cond} (or \ref{ex:cond:f}) elucidates the \textit{flow-sensitive} approach of the analysis: the resulting Alias Diagram (figure \ref{ex:cond:e} or \ref{ex:cond:f}) reports that either $x$ may be aliased to $a$ or may be aliased to $b$ but not both. Furthermore, the diagrams also report that $a$ may not be aliased to $b$ as a result of executing the instruction.

\subsubsection{Loops. } \eifkw{loop} \eif{ p } \eifkw{end} is the instruction that executes $p$ any number of times including none. \texttt{AC-Loop} captures this semantics by unioning $i$ times $G \alias p$, so it can produce $G$ (when $i=0$), or  $((G \alias p) \alias p)$ (when $i=2$), or $(((G \alias p) \alias p) \alias p)$ (when $i=3$) and so on. Figure \ref{ex:loop} shows an example on how the rule is applied to the instruction \eifkw{loop}\eif{ l := l.right}\eifkw{ end} on the Alias Diagram $G$ in figure \ref{ex:loop:a}. Consider a common example where \eif{l} is a linked list that contains a reference to its \eif{right} element. Figure \ref{ex:loop:b} shows the result of applying $G \alias $\eif{ l := l.right} and figure \ref{ex:loop:c} the result of $(G \alias $\eif{ l := l.right}$) \alias $\eif{ l := l.right}. Figure \ref{ex:loop:d} shows the final result:  $\bigcup\limits_{i \in \mathbb{N}}$ (\eif{$G_i \alias $ l := l.right}).

\begin{figure}[h!]
\centering
\begin{subfigure}[b]{0.4\textwidth}
    \begin {tikzpicture}[-latex ,auto ,node distance =2 cm  ,on grid ,
semithick , state/.style ={ circle ,top color =white ,  draw, minimum width =0.6 cm}]

    \node[state] (A) {\underline{$n_0$}};
    \node[state] (B) [right =of A] {$n_1$};
    \node[state] (C) [right =of B] {$n_2$};
    \node[state] (D) [right =of C] {$n_3$};
    \path (A) edge [] node {$l$} (B);
    \path (B) edge [] node {$right$} (C);
    \path (C) edge [] node {$right$} (D);
\end{tikzpicture}
\caption{}
\label{ex:loop:a}
\end{subfigure}

\begin{subfigure}[b]{0.4\textwidth}
    \begin {tikzpicture}[-latex ,auto ,node distance =2 cm  ,on grid ,
semithick , state/.style ={ circle ,top color =white ,  draw, minimum width =0.6 cm}]

    \node[state] (A) {\underline{$n_0$}};
    \node[state] (C) [right =4cm of A] {$n_2$};
    \node[state] (D) [right =of C] {$n_3$};
    \path (A) edge [] node {$l$} (C);
    \path (C) edge [] node {$right$} (D);
\end{tikzpicture}
\caption{}
\label{ex:loop:b}
\end{subfigure}
    
\begin{subfigure}[b]{0.4\textwidth}
    \begin {tikzpicture}[-latex ,auto ,node distance =2 cm  ,on grid ,
semithick , state/.style ={ circle ,top color =white ,  draw, minimum width =0.6 cm}]

    \node[state] (A) {\underline{$n_0$}};
    \node[state] (D) [right =6cm of A] {$n_3$};
    \path (A) edge [] node {$l$} (D);
\end{tikzpicture}
\caption{}
\label{ex:loop:c}
\end{subfigure}



\begin{subfigure}[b]{0.4\textwidth}
    \begin {tikzpicture}[-latex ,auto ,node distance =2 cm  ,on grid ,
semithick , state/.style ={ circle ,top color =white ,  draw, minimum width =0.6 cm}]

    \node[state] (A) {\underline{$n_0$}};
    \node[state] (B) [right =of A] {$n_1$};
    \node[state] (C) [right =of B] {$n_2$};
    \node[state] (D) [right =of C] {$n_3$};
    \path (A) edge [] node {$l$} (B);
    \path (A) edge [bend left] node {$l$} (C);
    \path (A) edge [bend left] node {$l$} (D);
    \path (B) edge [] node {$right$} (C);
    \path (C) edge [] node {$right$} (D);
\end{tikzpicture}
\caption{}
\label{ex:loop:d}
\end{subfigure}
\caption{Applying rule \texttt{AC-Loop}}
\label{ex:loop}
\end{figure}

Rule \texttt{AC-Loop} introduces imprecision but retains soundness. An optimization of the rules is to consider the loop condition. In the general case, determining loop termination is undecidable, but there are specific cases that can be asserted, e.g. the approach might be able to determine whether two variables \eif{v} and \eif{w} are already aliased, as in \eifkw{until} \eif{ v = w}\eifkw{ loop}\eif{ p}\eifkw{ end} (same concept can be applied to condition in rule \texttt{AC-Cond}).

\subsubsection{Unqualified Calls. } In rule \texttt{AC-UQCall}, \eif{l} and \eif{$f^\bullet$} are the lists of actual and formal arguments of routine \eif{f}, respectively. \eif{$\mid f\mid$} its body. The rule deals with unqualified calls (calls to routines of the Current object). Figure \ref{ex:unq} shows an example on how the rule is applied to the instruction \eif{set\_x (a)} on the Alias Diagram $G$ in figure \ref{ex:unq:a}. \eif{set\_x} is a routine defined as \eif{set\_x (v: T)}\eifkw{ do} \eif{x := v}\eifkw{ end}, it receives an argument \eif{v} of any arbitrary type \eif{T} and assigns it to variable \eif{x}. The semantics for \eif{$G \alias $} ($call$\eif{ set\_x (a)}) is 

$$
\overbrace{\underbrace{G [[v]:[a]]}_\text{see figure \ref{ex:unq:b}} \alias \eif{$\mid$ x := v $\mid$}}^\text{see figure \ref{ex:unq:c}}
$$

\begin{figure}[h!]
\centering
\begin{subfigure}[b]{0.3\textwidth}
    \begin {tikzpicture}[-latex ,auto ,node distance =2 cm  ,on grid ,
semithick , state/.style ={ circle ,top color =white ,  draw, minimum width =0.6 cm}]

    \node[state] (A)                    {\underline{$n_0$}};
    \node[state] (B) [right =of A] {$n_1$};
    \node[state] (C) [below =1.5cm of A] {$n_2$};
    \node[state] (D) [right =of C] {$n_3$};
    \path (A) edge [] node {$a$} (B);
    \path (A) edge [] node {$b$} (C);
    \path (A) edge [below] node {$x$} (D);
    
\end{tikzpicture}
\caption{}
\label{ex:unq:a}
\end{subfigure}
    \hfill 
\begin{subfigure}[b]{0.3\textwidth}
    \begin {tikzpicture}[-latex ,auto ,node distance =2 cm  ,on grid ,
semithick , state/.style ={ circle ,top color =white ,  draw, minimum width =0.6 cm}]

    \node[state] (A)                    {\underline{$n_0$}};
    \node[state] (B) [right =of A] {$n_1$};
    \node[state] (C) [below =1.5cm of A] {$n_2$};
    \node[state] (D) [right =of C] {$n_3$};
    \path (A) edge [] node {$a$} (B);
    \path (A) edge [bend right, dashed] node {$v$} (B);
    \path (A) edge [] node {$b$} (C);
    \path (A) edge [below] node {$x$} (D);
    
\end{tikzpicture}
\caption{}
\label{ex:unq:b}
\end{subfigure}
    \hfill 
\begin{subfigure}[b]{0.3\textwidth}
    \begin {tikzpicture}[-latex ,auto ,node distance =2 cm  ,on grid ,
semithick , state/.style ={ circle ,top color =white ,  draw, minimum width =0.6 cm}]

    \node[state] (A)                    {\underline{$n_0$}};
    \node[state] (B) [right =of A] {$n_1$};
    \node[state] (C) [below =1.5cm of A] {$n_2$};
    
    \path (A) edge [] node {$a$} (B);
    \path (A) edge [bend right, dashed] node {$v$} (B);
    \path (A) edge [] node {$b$} (C);
    \path (A) edge [bend left] node {$x$} (B);
    
\end{tikzpicture}
\caption{}
\label{ex:unq:c}
\end{subfigure}
\caption{Applying rule \texttt{AC-UQCall}}
\label{ex:unq}
\end{figure}

\begin{figure}[h!]
\centering
\begin{subfigure}[b]{0.3\textwidth}
    \begin {tikzpicture}[-latex ,auto ,node distance =2 cm  ,on grid ,
semithick , state/.style ={ circle ,top color =white ,  draw, minimum width =0.6 cm}]

    \node[state] (A)                    {\underline{$n_0$}};
    \node[state] (B) [right =of A] {$n_1$};
    \node[state] (C) [below =1.5cm of A] {$n_2$};
    \node[state] (D) [right =of C] {$n_3$};
    \path (A) edge [] node {$a$} (B);
    \path (A) edge [] node {$b$} (C);
    \path (A) edge [below] node {$x$} (D);
    
\end{tikzpicture}
\caption{$G$}
\label{ex2:unq:a}
\end{subfigure}
    \hfill 
\begin{subfigure}[b]{0.3\textwidth}
    \begin {tikzpicture}[-latex ,auto ,node distance =2 cm  ,on grid ,
semithick , state/.style ={ circle ,top color =white ,  draw, minimum width =0.6 cm}]

    \node[state] (A)                    {\underline{$n_0$}};
    \node[state] (B) [right =of A] {$n_1$};
    \node[state] (C) [below =1.5cm of A] {$n_2$};
    
    \path (A) edge [] node {$a$} (B);
    \path (A) edge [] node {$b$} (C);
    \path (A) edge [bend left] node {$x$} (B);
    
\end{tikzpicture}
\caption{$G \alias set\_x (a)$}
\label{ex2:unq:b}
\end{subfigure}
    \hfill 
\begin{subfigure}[b]{0.3\textwidth}
    \begin {tikzpicture}[-latex ,auto ,node distance =2 cm  ,on grid ,
semithick , state/.style ={ circle ,top color =white ,  draw, minimum width =0.6 cm}]

    \node[state] (A)                    {\underline{$n_0$}};
    \node[state] (B) [right =of A] {$n_1$};
    \node[state] (C) [below =1.5cm of A] {$n_2$};
    
    \path (A) edge [] node {$a$} (B);
    \path (A) edge [] node {$b$} (C); 
    \path (A) edge [bend left] node {$x$} (C);
    
\end{tikzpicture}
\caption{$G \alias (set\_x (a); set\_x (b))$}
\label{ex2:unq:c}
\end{subfigure}
\caption{Example of call-site-sensitivity.}
\label{ex2:unq}
\end{figure}
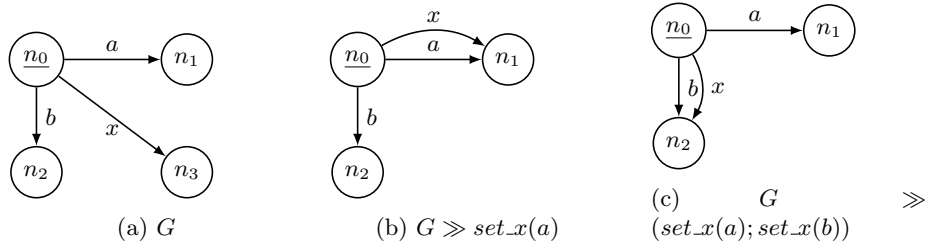

The analysis is context-sensitive, meaning that it differentiates between executions of a given instruction in different contexts. In particular, the analysis is call-site-sensitive, meaning that it does not coalesce the effects of different calls to the same routine. Figure \ref{ex2:unq} depicts the process of performing $G \alias ($\eifkw{call }\eif{ set\_x (a);}\eifkw{ call }\eif{ set\_x (b))}. A context-insensitive analysis could deduce that this may alias both $a$ and $b$ to $x$ and hence (wrongly) to each other. As shown in figure \ref{ex2:unq:c} only $b$ is aliased to $x$.

\subsubsection{Qualified calls:} Rule \texttt{AC-QCall} deals with qualified calls (calls to routines on a different object from Current). Figure \ref{ex:q} shows an example on how the rule is applied to the instruction \eif{a.set\_x (b)} on the Alias Diagram $G$ in figure \ref{ex:q:a}. The semantics of \eif{a.set\_x (b)} is apply routine \eif{set\_x} on the object attached to \eif{a} (routine \eif{set\_x} has the same definition as before). The semantics for \eif{$G \alias $} (\eif{a.}\eifkw{call}\eif{ set\_x (b)}) is 

$$\overbrace{\eifkw{reroot}_{
        \underbrace{(
        	\underbrace{\eifkw{reroot}_{a'\bullet G} V(a)}_\text{see figure \ref{ex:q:b}}
            \alias 
            \eifkw{call}  set\_x(a'.b)
        )}_\text{see figure \ref{ex:q:c}}} R_G}^\text{see figure \ref{ex:q:d}}
        $$

Figure \ref{ex:q:b} performs dot distribution over $G$ (see section \ref{sec:dot-dist}) and re-roots the graph to $V (a)$ (the set of nodes reachable from the root through $a$, in this case $\{n_1\}$). Figure \ref{ex:q:c} depicts the resulting Alias Diagram after applying rule \texttt{AC-UQCall} on the Alias Diagram in figure \ref{ex:q:b}. Figure \ref{ex:q:c} shows the importance of shifting the context: operations are applied to $G$ with $n_1$ (the object attached to $a$) as the root. Finally, figure \ref{ex:q:d} re-roots the graph to its initial roots. This operation allows the analysis to support any number of nested calls. Each qualified call will \textit{(i)} perform dot distribution, allowing the analysis to have access to those variables of the source (e.g. variable \eif{b} when analysing the call \eif{a.set\_x (b)}); \textit{(ii)} shift the context (as in an O-O computation), allowing the analysis to perform operations on the right object; \textit{(iii)} shift the roots back, allowing the analysis to continue the normal operation.

\begin{figure}[h!]
\centering
\begin{subfigure}[b]{0.3\textwidth}
    \begin {tikzpicture}[-latex ,auto ,node distance =2 cm  ,on grid ,
semithick , state/.style ={ circle ,top color =white ,  draw, minimum width =0.6 cm}]

    \node[state] (A) {\underline{$n_0$}};
    \node[state] (B) [right =of A] {$n_1$};
    \node[state] (D) [below =1.3cm of A] {$n_3$};
    \node[state] (C) [right =of D] {$n_2$};
    \node[state] (E) [right =of B] {$n_4$};

    \path (A) edge [] node {$a$} (B);
    \path (B) edge [] node {$x$} (E);
    \path (A) edge [] node {$b$} (C);
    \path (A) edge [] node {$x$} (D);

\end{tikzpicture}
\caption{}
\label{ex:q:a}
\end{subfigure}
    \hfill 
\begin{subfigure}[b]{0.3\textwidth}
    \begin {tikzpicture}[-latex ,auto ,node distance =2 cm  ,on grid ,
semithick , state/.style ={ circle ,top color =white ,  draw, minimum width =0.6 cm}]

    \node[state] (A) {$n_0$};
    \node[state] (B) [right =of A] {\underline{$n_1$}}; 
    \node[state] (D) [below =1.3cm of A] {$n_3$};
    \node[state] (C) [right =of D] {$n_2$};
    \node[state] (E) [right =of B] {$n_4$};

    \path (A) edge [] node {$a$} (B);
    \path (B) edge [] node {$x$} (E);
    \path (A) edge [] node {$b$} (C);
    \path (A) edge [] node {$x$} (D);
    \path (B) edge [bend right, dashed, above] node {$a'$} (A);
    
\end{tikzpicture}
\caption{}
\label{ex:q:b}
\end{subfigure}
    
\begin{subfigure}[b]{0.3\textwidth}
    \begin {tikzpicture}[-latex ,auto ,node distance =2 cm  ,on grid ,
semithick , state/.style ={ circle ,top color =white ,  draw, minimum width =0.6 cm}]

    \node[state] (A) {$n_0$};
    \node[state] (B) [right =of A] {\underline{$n_1$}}; 
    \node[state] (D) [below =1.3cm of A] {$n_3$};
    \node[state] (C) [right =of D] {$n_2$};

    \path (A) edge [] node {$a$} (B);
    \path (A) edge [] node {$b$} (C);
    \path (A) edge [] node {$x$} (D);
    \path (B) edge [bend right, dashed, above] node {$a'$} (A);
    \path (A) edge [bend right, dashed] node {$v$} (C);
    \path (B) edge [] node {$x$} (C);
\end{tikzpicture}
\caption{}
\label{ex:q:c}
\end{subfigure}
\hfill 
\begin{subfigure}[b]{0.3\textwidth}
    \begin {tikzpicture}[-latex ,auto ,node distance =3 cm  ,on grid ,
semithick , state/.style ={ circle ,top color =white ,  draw, minimum width =0.6 cm}]

    \node[state] (A) {\underline{$n_0$}};
    \node[state] (B) [right =of A] {$n_1$};
    \node[state] (D) [below =1.3cm of A] {$n_3$};
    \node[state] (C) [right =of D] {$n_2$};

    \path (A) edge [] node {$a$} (B);
    \path (B) edge [bend right, dashed, above] node {$a'$} (A);
    \path (A) edge [bend right, dashed] node {$v$} (C);
    \path (A) edge [] node {$b$} (C);
    \path (A) edge [] node {$x$} (D);
    \path (B) edge [] node {$x$} (C);
    
\end{tikzpicture}
\caption{}
\label{ex:q:d}
\end{subfigure}
\caption{Applying rule \texttt{AC-QCall}}
\label{ex:q}
\end{figure}

\section{AutoAlias: a graph-based implementation for the Alias calculus}
\label{section:autoalias}
AutoAlias is a graph-based implementation for the Alias Calculus, sources of the tool are available in \cite{AutoAlias:Impl} and results can be checked in \cite{AutoAliasAutoFrame:results}.  

One of the main concerns of a graph-based approach w.r.t. to the relation-based one (an approach from \cite{Kogtenkov:2015, Meyer:Alias:14, BM:2010}) is the performance of the computation, especially when dealing with conditionals and loops (including recursion). This can be seen in rules \texttt{AC-Cond} and \texttt{AC-Loop} from section \ref{alias-calc}: both rules perform union operations on graphs. Sections \ref{handling-cond} and \ref{handling-loops} show the techniques being used when dealing with such cases. Section \ref{handling-dynbin} explains how Dynamic Binding (and Inheritance and Polymorphism), an important property of O-O computations is being handled. 

\subsection{Handling conditionals}
\label{handling-cond}
The non-deterministic choice instruction has the form

\noindent
\eifkw{then}\\
\hspace*{.5cm}$branch_1$\\
\eifkw{elseif}\\
\hspace*{.5cm}$branch_2$\\
\ldots\\
\eifkw{else}\\
\hspace*{.5cm}$branch_{n}$\\
\eifkw{end}

According to the \texttt{AC-Cond} rule, each branch of the conditional ($branch_1$ \ldots $branch_n$) is analyzed with a starting Alias Diagram $G$ that holds initially. Then, the resulting diagrams are cloned and union. When processing $branch_i$, where $i \in 1 \ldots n$, the implementation maintains two sets: $A_i \in T \rightarrow O \rightarrow O$ (insertions -- $A$ for Additions -- of references in the alias diagram) and $D_i \in T \rightarrow O \rightarrow O$ (deletions -- $D$ for Deletions -- of references). Both sets will contain triples $(name, source, target)$.
 
At the end of processing branch $i$, the implementation removes all the elements of $A_i$ and add all elements of $D_i$ to the Alias Graph (so as to get back to the starting state).

At the end of processing the conditional, $branch_{n}$, the implementation:
\begin{enumerate}[label=(\roman*)]
    \item \eifkw{clone}s the root of the diagram  $n-1$ times;
    \item for all $b \in 2 \ldots n$ and for all $(n,s,t) \in D_b$, adds $(n, R^b, t)$ to the Alias Diagram;
    \item for all $b \in 2 \ldots n$ and for all $(n,s,t) \in A_b$, adds $(names (s, t), R^b, t)$ to the Alias Diagram, where $names (source, target)$ is a function returning the set of names from $source$ to $target$ in the Alias Diagram;
    \item changes the corresponding clone root in sets $A$ and $D$;
    \item inserts the union of $A_i$ and removes the union of $D_i$ in the Alias Diagram.
\end{enumerate} 

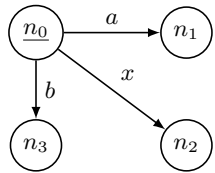
\begin{figure}[h!]
\centering
\begin {tikzpicture}[-latex ,auto ,node distance =2 cm  ,on grid ,
semithick , state/.style ={ circle ,top color =white ,  draw}]

        \node[state] (n0)           {\underline{$n_0$}};
        \node[state] (n1) [right =of n0] {$n_1$};
        \node[state] (n2) [below =1.5cm of n1] {$n_2$};
        \node[state] (n3) [below =1.5cm of n0] {$n_3$};
        
        \path  
            (n0) edge [] node {$a$} (n1)
            (n0) edge [] node {$x$} (n2)
            (n0) edge [] node {$b$} (n3);
    \end{tikzpicture}
\caption{Alias Diagram $G$}
\label{ex:handling-cond}
\end{figure}

Consider the Alias Diagram in figure \ref{ex:handling-cond}. The inital process of applying \eif{$G \alias $} (\eifkw{then}\eif{ a := x}\eifkw{ else}\eif{ b := x}\eifkw{ end}) is depicted in figure \ref{ex:handling-cond2}.

Figure \ref{ex:hand-cond:a} applies $G \alias $\eif{ a := x}. The implementation maintains sets $A$ and $D$:
\begin{align*} 
A= &  1: \{(a, \text{\underline{$n_0$}}, n_2)\} \\
D= &  1: \{(a, \text{\underline{$n_0$}}, n_1)\} \\ 
\end{align*}

Figure \ref{ex:hand-cond:b} is the result of removing all elements of $A_{branch_1}$ and adding all elements of  $D_{branch_1}$ (to get back to the starting state). Figure \ref{ex:hand-cond:c} applies $G \alias $\eif{ b := x}, maintaining sets $A$ and $D$:

\begin{align*} 
A= &  1: \{(a, \text{\underline{$n_0$}}, n_2)\} \\
&  2: \{(b, \text{\underline{$n_0$}}, n_2)\} \\
D= &  1: \{(a, \text{\underline{$n_0$}}, n_1)\} \\ 
&  2: \{(b, \text{\underline{$n_0$}}, n_3)\} \\
\end{align*}

\begin{figure}[h!]
\centering
\begin{subfigure}[b]{0.3\textwidth}
    \begin {tikzpicture}[-latex ,auto ,node distance =2 cm  ,on grid ,
semithick , state/.style ={ circle ,top color =white ,  draw, minimum width =0.6 cm}]

    \node[state] (n0)           {\underline{$n_0$}};
        \node[state] (n2) [below =1.3cm of n1] {$n_2$};
        \node[state] (n3) [below =1.3cm of n0] {$n_3$};
        
        \path  
            (n0) edge [bend left] node {$a$} (n2)
            (n0) edge [] node {$x$} (n2)
            (n0) edge [] node {$b$} (n3);
    \end{tikzpicture}
\caption{}
\label{ex:hand-cond:a}
\end{subfigure}
    \hfill 
\begin{subfigure}[b]{0.3\textwidth}
    \begin {tikzpicture}[-latex ,auto ,node distance =2 cm  ,on grid ,
semithick , state/.style ={ circle ,top color =white ,  draw, minimum width =0.6 cm}]

    \node[state] (n0)           {\underline{$n_0$}};
        \node[state] (n1) [right =of n0] {$n_1$};
        \node[state] (n2) [below =1.3cm of n1] {$n_2$};
        \node[state] (n3) [below =1.3cm of n0] {$n_3$};
        
        \path  
            (n0) edge [] node {$a$} (n1)
            (n0) edge [] node {$x$} (n2)
            (n0) edge [] node {$b$} (n3);
    \end{tikzpicture}
\caption{}
\label{ex:hand-cond:b}
\end{subfigure}
    \hfill 
\begin{subfigure}[b]{0.3\textwidth}
    \begin {tikzpicture}[-latex ,auto ,node distance =2 cm  ,on grid ,
semithick , state/.style ={ circle ,top color =white ,  draw, minimum width =0.6 cm}]

    \node[state] (A)                    {\underline{$n_0$}};
    \node[state] (B) [right =of A] {$n_1$};
    \node[state] (D) [below =1.3cm of B] {$n_2$};
    \path (A) edge [] node {$a$} (B);
    \path (A) edge [bend right, left] node {$b$} (D);
    \path (A) edge [] node {$x$} (D);

\end{tikzpicture}
\caption{}
\label{ex:hand-cond:c}
\end{subfigure}
\caption{Applying rule \texttt{AC-Cond}}
\label{ex:handling-cond2}
\end{figure}

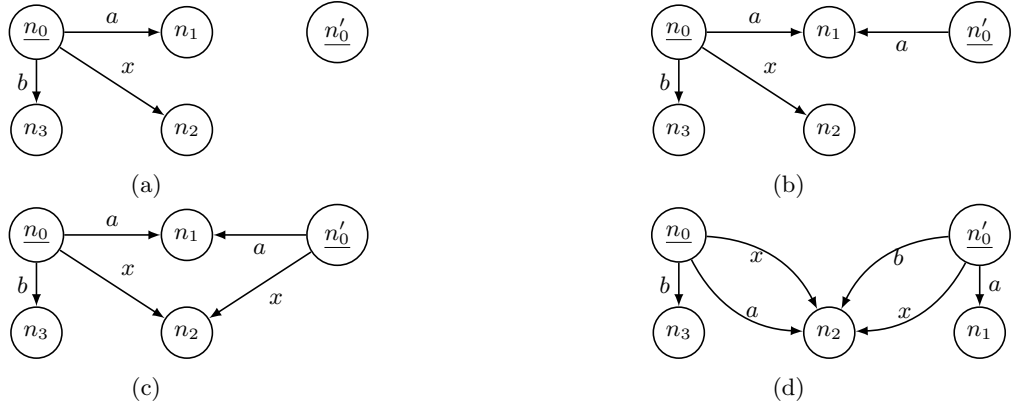
\begin{figure}[h!]
\centering
\begin{subfigure}[b]{0.3\textwidth}
    \begin {tikzpicture}[-latex ,auto ,node distance =2 cm  ,on grid ,
semithick , state/.style ={ circle ,top color =white ,  draw, minimum width =0.6 cm}]

     \node[state] (n0)                    {\underline{$n_0$}};
        \node[state] (n3) [below =1.3cm of n0] {$n_3$};
        \node[state] (n2) [right =of  n3] {$n_2$};
        \node[state] (n1) [right =of  n0] {$n_1$};
        \node[state] (n0p) [right =of  n1] {\underline{$n_0'$}};

        \path  
            (n0) edge [left] node {$b$} (n3)
            (n0) edge [] node {$a$} (n1)
            (n0) edge [] node {$x$} (n2);
    \end{tikzpicture}
\caption{}
\label{ex:handling-cond-end-a}
\end{subfigure}
    \hfill 
\begin{subfigure}[b]{0.3\textwidth}
    \begin {tikzpicture}[-latex ,auto ,node distance =2 cm  ,on grid ,
semithick , state/.style ={ circle ,top color =white ,  draw, minimum width =0.6 cm}]

     \node[state] (n0)                    {\underline{$n_0$}};
        \node[state] (n3) [below =1.3cm of n0] {$n_3$};
        \node[state] (n2) [right =of  n3] {$n_2$};
        \node[state] (n1) [right =of  n0] {$n_1$};
        \node[state] (n0p) [right =of  n1] {\underline{$n_0'$}};

        \path  
            (n0) edge [left] node {$b$} (n3)
            (n0) edge [] node {$a$} (n1)
            (n0) edge [] node {$x$} (n2);
        \path (n0p) edge [] node {$a$} (n1);
    \end{tikzpicture}
\caption{}
\label{ex:handling-cond-end-b}
\end{subfigure}

    \begin{subfigure}[b]{0.3\textwidth}
    \begin {tikzpicture}[-latex ,auto ,node distance =2 cm  ,on grid ,
semithick , state/.style ={ circle ,top color =white ,  draw, minimum width =0.6 cm}]

    \node[state] (n0)                    {\underline{$n_0$}};
        \node[state] (n3) [below =1.3cm of n0] {$n_3$};
        \node[state] (n2) [right =of  n3] {$n_2$};
        \node[state] (n1) [right =of  n0] {$n_1$};
        \node[state] (n0p) [right =of  n1] {\underline{$n_0'$}};

        \path  
            (n0) edge [left] node {$b$} (n3)
            (n0) edge [] node {$a$} (n1)
            (n0) edge [] node {$x$} (n2);
        \path (n0p) edge [] node {$a$} (n1)
            (n0p) edge [] node {$x$} (n2);
    \end{tikzpicture}
\caption{}
\label{ex:handling-cond-end-c}
\end{subfigure}
    \hfill 
\begin{subfigure}[b]{0.3\textwidth}
    \begin {tikzpicture}[-latex ,auto ,node distance =2 cm  ,on grid ,
semithick , state/.style ={ circle ,top color =white ,  draw, minimum width =0.6 cm}]

    \node[state] (n0)                    {\underline{$n_0$}};
        \node[state] (n3) [below =1.3cm of n0] {$n_3$};
        \node[state] (n2) [right =of  n3] {$n_2$};
        \node[state] (n1) [right =of  n2] {$n_1$};
        \node[state] (n0p) [above =1.3cm of  n1] {\underline{$n_0'$}};

        \path  
            (n0) edge [left] node {$b$} (n3)
            (n0) edge [bend left, left] node {$x$} (n2)
            (n0) edge [bend right, right] node {$a$} (n2);
        \path (n0p) edge [] node {$a$} (n1)
            (n0p) edge [bend left, left] node {$x$} (n2)
            (n0p) edge [bend right, right] node {$b$} (n2);
    \end{tikzpicture}
\caption{}
\label{ex:handling-cond-end-d}
\end{subfigure}
\caption{Process at the end of analyzing a conditional}
\label{ex:handling-cond-end}
\end{figure}

The Alias Diagram is restored by removing all elements of $A_{branch_2}$ and adding all elements of  $D_{branch_2}$ (to get back to the starting state -- as depicted in figure \ref{ex:handling-cond}). At the end of processing the conditional, the implementation: \textit{(i)} \eifkw{clone}s the root of the diagram  $n-1$ times. In this case, only once, as shown in figure \ref{ex:handling-cond-end-a}; then, \textit{(ii)} for all $b \in 2 \ldots n$ and for all $(n,s,t) \in D_b$, adds $(n, R^b, t)$ to the Alias Diagram, see figure \ref{ex:handling-cond-end-b}; the implementation then \textit{(iii)} for all $b \in 2 \ldots n$ and for all $(n,s,t) \in A_b$, adds $(names (s, t), R^b, t)$ to the Alias Diagram, as depicted in figure \ref{ex:handling-cond-end-c}; \textit{(iv)} changes the corresponding clone root in sets $A$ and $D$
\begin{align*} 
A= &  1: \{(\text{\underline{$n_0$}}, a, \{n_2\})\} \\
&  2: \{(\text{\underline{$n_0'$}}, b, \{n_2\})\} \\
D= &  1: \{(\text{\underline{$n_0$}}, a, \{n_1\})\} \\ 
&  2: \{(\text{\underline{$n_0'$}}, b, \{n_3\})\} \\
\end{align*}
; and finally, \textit{(v)} the resulting Alias Diagram is the result of inserting the union of $A_i$ and removing the union of the $D_i$ (figure \ref{ex:handling-cond-end-d}).

The process is an optimization of the rules. Notice that each branch of a conditional will only change a small part of the diagram, this is being handled by just copying the edges that are being modified by the program.


\subsection{Handling loops}
\label{handling-loops}
Consider the instruction $G \alias $ \eifkw{loop} \eif{ p } \eifkw{end} as 
 executing the instruction \eif{p} any number of times ($i$) including none, and unioning the resulting Alias Diagrams so it can produce $G$ (when $i=0$), or  $((G \alias p) \alias p)$ (when $i=2$), or $(((G \alias p) \alias p) \alias p)$ (when $i=3$) and so on.
 
Then the mechanism to handle loops gives the following process:
\begin{itemize}
\item Use a single $D$ (deletion) set. Here there is no need for $A$ sets.
\item Process the loop body (\eif{p}) repeatedly, at each iteration adding deleted references to $D$.
\item At the end of each iteration, nothing special needs to be done.
\item Stop when reaching a fixpoint.
\item At the end of the process, re-insert the elements of $D$.
\end{itemize}

The general idea can be applied to recursion. 

\begin{itemize}
\item Maintain a single $D$ (deletion) set, as well as a stack with each call (and its target object).
\item For each call, update the stack (so to handle the different ways of recursion, e.g. direct or indirect recursion).
\item Process the feature body repeatedly, at each call adding deleted references to $D$.
\item Stop when reaching a fixpoint by using the stack calls.
\item At the end of the process, re-insert the elements of $D$.
\end{itemize}

\paragraph{Termination:} Termination of fixpoint computations.
 
\begin{lemma}
if the analysis starts from an existing graph and the program does not perform any object creations, then the iteration process (for loops) reaches a fixpoint finitely.
\end{lemma}

Proof of the lemma: the graph is finite; each iteration does not remove any nodes or edges, and can only insert edges. This cannot go on forever.

Hence the process to guarantee termination, as follows. It assumes that we associate with every creation instruction \eifkw{create }\eif{X} a positive integer $N$ (in the simplest variant, N = 1) and a fresh variable \eif{fx}.

\begin{tabular}{lp{10cm}}
S1. & The first $N$ times processing the instruction, apply the normal rule (remove all edges labeled \eif{x} from the root, create new node, create edge labeled \eif{x} from the root to that node).\\
 
S2.& The $N$-th time processing the instruction, after doing S1, add the label \eif{fx} to the new edge (i.e. alias \eif{fx} to \eif{x}). 
 \\
S3.& Every subsequent time processing the instruction (starting with the $N+$1st), treat it not through the creation instruction rule but as if it were the assignment \eif{x := fx}.
\end{tabular}
  
With the policy, after some number of iterations no new node will ever be created. So the lemma applies and the fixpoint process terminates.

\subsection{Dynamic Binding}
\label{handling-dynbin} 
One of the main mechanisms of O-O programming is inheritance. It enables users to create `is-a' relations between different classes: considering \eif{A} and \eif{B} as types, if \eif{B} inherits from \eif{A}, whenever an instance of \eif{A} is required, an instance of B will be acceptable. This mechanism enables entities to be polymorphic: an entity is polymorphic if at run-time, its type differs from its static type.

Dynamic binding is the property that any execution of a feature call will use the version of the feature best adapted to the type of the target object, versions might differ thanks to Polymorphism. It is important to mention how AutoAlias handles this property since AutoAlias statically analyze the source code, hence it is not possible to determine what is the appropriate type of a specific entity. Consider, as an example, the classes depicted in figure \ref{ex:dynBin}. Class \eif{T1}, in figure \ref{ex:dynBin:a}, defines two variables \eif{c} and \eif{b}. Class \eif{T2}, in figure \ref{ex:dynBin:b}, inherits from class \eif{T1} (by using the keyword \eifkw{inherit}), it also gives a redefinition of routine \eif{set} (indicated by the keyword \eifkw{redefine}). 

\begin{figure}[h!]
    \centering
    \begin{subfigure}[b]{0.3\textwidth}
    \[
        \begin{array}{l}
    	\eifkw{class}\eif{ T1}\\
		\eifkw{feature}\\
        \hspace*{.5cm}\eif{c, b: T}\\
        \hspace*{.5cm}\eif{set (arg: T) }\eifkw{ do}\\
		\hspace*{1cm}\eif{ c := arg}\\
		\hspace*{.5cm}\eifkw{end}\\
        \eifkw{end}\\
        
        \end{array}
    \]
    \caption{}
    \label{ex:dynBin:a}
    \end{subfigure}
    \hfill 
    \begin{subfigure}[b]{0.5\textwidth}
    \[
        \begin{array}{l}
    	\eifkw{class}\eif{ T2}\\
    	\eifkw{inherit}\eif{ T1 }\eifkw{ redefine }\eif{ set }\eifkw{ end}\\
		\eifkw{feature}\\
        \hspace*{.5cm}\eif{set (arg: T) }\eifkw{ do}\\
		\hspace*{1cm}\eif{ b := arg}\\
		\hspace*{.5cm}\eifkw{end}\\
        \eifkw{end}\\
        
        \end{array}
      \]
    \caption{}
    \label{ex:dynBin:b}
    \end{subfigure}
    
    \begin{subfigure}[b]{0.5\textwidth}
    \[
        \begin{array}{l}
    	\eifkw{class}\eif{ B}\\
    	\eifkw{feature}\\
    	\hspace*{.5cm}\eif{t, a: T1}\\
        \hspace*{.5cm}\eif{call\_set }\eifkw{ do}\\
		\hspace*{1cm}\eif{ t.set (a)}\\
		\hspace*{.5cm}\eifkw{end}\\
        \eifkw{end}\\
        
        \end{array}
      \]
    \caption{}
    \label{ex:dynBin:call}
    \end{subfigure}
    \caption{Dynamic Binding example}
    \label{ex:dynBin}
\end{figure}

A must-aliasing approach will yield, after executing feature \eif{call\_set} in figure \ref{ex:dynBin:call}, a result that depends on the dynamic type of the entity \eif{t}: if during execution \eif{t} is attached to an object of type \eif{T1}, the result would be  \eif{t.c} is aliased to \eif{a};  if during execution \eif{t} is attached to an object of type \eif{T2}, the result would be \eif{t.b} is aliased to \eif{a}.

A may-aliasing approach (as the one adopted by AutoAlias) would yield that \eif{t.c} may be aliased to \eif{a} or that \eif{t.b} may be aliased to \eif{a}. The mechanism implemented by AutoAlias is to treat the instruction as a conditional, in this case it would be:

\noindent
\eifkw{then}\\
\hspace*{.5cm}t.set (a) \textit{ -- considering \eif{t} attached to \eif{T1}}\\
\eifkw{else}\\
\hspace*{.5cm}t.set (a) \textit{ -- considering \eif{t} attached to \eif{T2}}\\
\eifkw{end}

\begin{figure}[h!]
\centering
\begin {tikzpicture}[-latex ,auto ,node distance =2 cm  ,on grid ,
semithick , state/.style ={ circle ,top color =white ,  draw}]

        \node[state] (n0)           {\underline{$n_0$}};
        \node[state] (n1) [right =of n0] {$n_1$};
        \node[state] (n3) [right =of n1] {$n_3$};
        \node[state] (n4) [below =1.5cm of n1] {$n_4$};
        \node[state] (n1p) [below =1.5cm of n4] {$n_1'$};
        \node[state] (n0p) [left =of n1p] {\underline{$n_0'$}};
        \node[state] (n2) [right =of n1p] {$n_2$};
        
        \path  
            (n0) edge [] node {$t$} (n1)
            (n0) edge [left] node {$a$} (n4);
        \path  
            (n1) edge [] node {$b$} (n3)
            (n1) edge [] node {$c$} (n4);
        \path  
            (n0p) edge [] node {$a$} (n4)
            (n0p) edge [] node {$t$} (n1p);
        \path  
            (n1p) edge [] node {$c$} (n2)
            (n1p) edge [right] node {$b$} (n4);
    \end{tikzpicture}
\caption{Alias Diagram after executing $G \alias ($ \eifkw{call }\eif{call\_set}$)$ from figure \ref{ex:dynBin:call}}
\label{ex:dynBin:AD}
\end{figure}
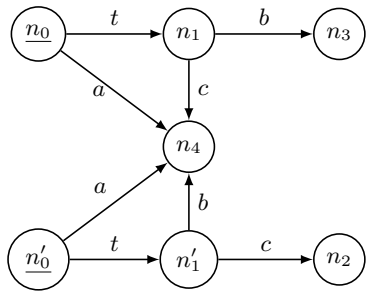

It will consider as many branches as heirs of \eif{T1} exists that redefine the feature call. For this particular case, the Alias Diagram resulting after executing $G \alias ($\eifkw{call }\eif{call\_set}$)$ is depicted in figure \ref{ex:dynBin:AD}.

The mechanism introduces imprecision but retains soundness. Notice that \eif{t.c} may be aliased to \eif{a} or \eif{t.b} may be aliased to \eif{a}, but \eif{t.b} may not be aliased to \eif{t.c}.

\subsection{Using AutoAlias}
\subsubsection{AutoFrame}
AutoFrame is a companion tool \cite{AutoFrame2019} that uses AutoAlias. AutoFrame produces the set of locations that are allowed to change in a routine. It statically analyzes the source code of a routine. AutoFrame relies on Autoalias to determine possibly aliasing. The most relevant results of AutoFrame so far  are \textit{(i)} the automatic reconstruction of the exact frame clauses, a total of 169 clauses, for an  8000+ lines data structures and algorithms. The frame inference in this case takes about 25 seconds on an ordinary laptop computer. \textit{(ii)} The automatic generation of frame conditions of a 150000 lines library for building GUIs. The frame inference in this case takes about 232 seconds. 

\subsubsection{Precision of AutoAlias}
Deutsch, in \cite{Deutsch:1994}, presents a comparison of the precision of some alias analysis algorithms (including his) on a structure-copying program creating two lists whose elements are pairwise aliased. The idea behind was to answer an open problem on how to improve the accuracy of alias analysis in the presence of recursive pointers data structure. We ran AutoAlias on this program to evaluate the accuracy of our approach. Figure \ref{alg:a} shows the algorithm used in \cite{Deutsch:1994}. Since AutoAlias receives as an input Eiffel code, Figure \ref{alg:b} shows the respective implementation of the algorithm.

\begin{figure}
    \centering
    \begin{subfigure}[b]{0.3\textwidth}
    \[
        \begin{array}{l}
    	\eif{struct List\{}\\
    	\hspace*{.5cm}\eif{char *hd;}\\
    	\hspace*{.5cm}\eif{struct List *tl;}\\
    	\eif{\};}\\ \\
    	\eif{struct List *}\\
        \eif{Copy(struct List *L) \{}\\
        \hspace*{.5cm}\eif{struct List *p, *t1;}\\
        \hspace*{.5cm}
        \eif{ if (L == }\eifkw{null}\eif{)}\\
        \hspace*{1cm}
        \eifkw{ return}\eif{(L);}\\
        \hspace*{.5cm}
        \eif{ p = malloc(\ldots);}\\
        \hspace*{.5cm}
        \eif{ tl = L$\rightarrow$ tl;}\\
        \hspace*{.5cm}
        \eif{ p$\rightarrow$tl = Copy(t1);}\\
        \hspace*{.5cm}
        \eif{ p$\rightarrow$hd = L$\rightarrow$hd;}\\
        \hspace*{.5cm}
        \eifkw{ return}\eif{(p);}\\
        \eif{\}}\\ \\
        \eif{/* X is an unliased list */}\\
        \texttt{$L_1:$ }\eif{t2 = X;}\\
        \texttt{$L_2:$ }\eif{Y = Copy (t2);}\\
        \texttt{$L_3:$ }\eif{X = }\eifkw{null;}\\
        
        \end{array}
    \]
    \caption{C-like}
    \label{alg:a}
    \end{subfigure}
    \hfill 
    \begin{subfigure}[b]{0.5\textwidth}
    \[
        \begin{array}{l}
    	\eifkw{class}\eif{ LST}\\
		\eifkw{feature}\\
        \eif{hd: T }\eifcomment{list head}
        \\
        \eif{tl: LST }\eifcomment{tail of the list}
        \\
      	\eifkw{end}
        \\ \\
        
        \eif{copy\_ (L: LST): LST}\\
		\hspace*{.5cm}\eifkw{local}\eif{ t1: LST}\\
		\hspace*{.5cm}\eifkw{do}\\
		\hspace*{1cm}\eifkw{if}\eif{ L $=$ }\eifkw{ Void~ then}\\
		\hspace*{1.5cm}\eifkw{create~ Result}\\
		\hspace*{1cm}\eifkw{else}\\
		\hspace*{1.5cm}\eifkw{create ~Result}\\
		\hspace*{1.5cm}\eif{t1 := L.tl}\\
		\hspace*{1.5cm}\eifkw{Result}\eif{.tl := copy\_ (t1)}\\
		\hspace*{1.5cm}\eifkw{Result}\eif{.hd := L.hd}\\
		\hspace*{1cm}\eifkw{end}\\
		\hspace*{.5cm}\eifkw{end}\\\\    
        \eifkw{local}\eif{ X,Y,t2: LST}\\
		\eifkw{do}\\
		\texttt{$L_0:$ }\hspace*{.5cm}\eif{create X}\\
		\texttt{$L_1:$ }\hspace*{.5cm}\eif{t2 := X}\\
		\texttt{$L_2:$ }\hspace*{.5cm}\eif{Y := copy\_ (t2)}\\
		\texttt{$L_3:$ }\hspace*{.5cm}\eif{create X}\\
		\eifkw{end}\\
        
        \end{array}
      \]
    \caption{Eiffel}
    \label{alg:b}
    \end{subfigure}
    \caption{Structure-copying program}
\end{figure}

Figure \ref{alg:a} defines the algorithm in a C-like program. \eif{List} is a structure containing two pointers: to the the head (\eif{char *hd}) and to the tail (\eif{List *tl}) of the list. \eif{Copy} is a procedure that returns a list which is a copy of the elements of the list being passed as an argument. In Figure \ref{alg:b}, the \eif{List} structure is implemented as a class (\eif{LST}) that contains two references: to the head (\eif{hd}) and to the tail (\eif{tl: LST}) of the list. The procedure \eif{copy\_} is an implementation of \eif{copy}. The return value of a routine in Eiffel is set by assigning it to the local variable \eifkw{Result}. The type of this local variable is the one defined as the return type in the signature of the routine (\eif{LST} in this case). Hence, there is not need to create the local variable \eif{p} as in Figure \ref{alg:a}, we directly use \eifkw{Result}. In Eiffel, class attributes have read-only privileges from outside the class: they can be changed only through procedures. This protects encapsulation and consistency of the object. Hence, the instruction \eifkw{Result}\eif{.tl := copy\_ (t1)} is not permitted unless the proper \eifkw{assigner} routines are being set. Figure \ref{alg:b} does not show the corresponding setter routines due to space.

\cite{Deutsch:1994} defines five program properties and compares the precision of five different algorithms (including theirs) for alias analysis. Table \ref{deutsch:results} shows that comparison, the table also adds the results by AutoAlias.

\begin{table}[t]
\hspace*{-1cm}
\centering
\begin{tabular}{p{5cm}llllll}
\textbf{Program Property} & \cite{Larus:1988} & \cite{Chase:1990} & \cite{Landi:1992} & \cite{Choi:1993} & \cite{Deutsch:1994} & AutoAlias \\
$P_1$: $X$ and $Y$ are acyclic                                         &           & yes       & yes       &           & yes       & yes         \\
$P_2$: two successive heads of $Y$ don't alias                         &           &           & yes       &           & yes       & yes         \\
$P_3$: $X$ and $Y$ tails don't alias                                   & yes       & yes       &           & yes       & yes       & yes         \\
$P_4$: heads of $X$ and $Y$ are aliased only pairwise                  &           &           &           &           & yes       & yes         \\
$P_5$: at point $L_3$, heads and tails of $Y$ are completely unaliased &           &           & yes       &           & yes       & yes         \\
                                                                       &           &           &           &           &           &            
\end{tabular}
\caption{Precision of alias analysis algorithms on a structure-copying program from figures \ref{alg:a} and \ref{alg:b} (adapted from \cite{Deutsch:1994})}
\label{deutsch:results}
\end{table}

\begin{figure}
\includegraphics[width=5in]{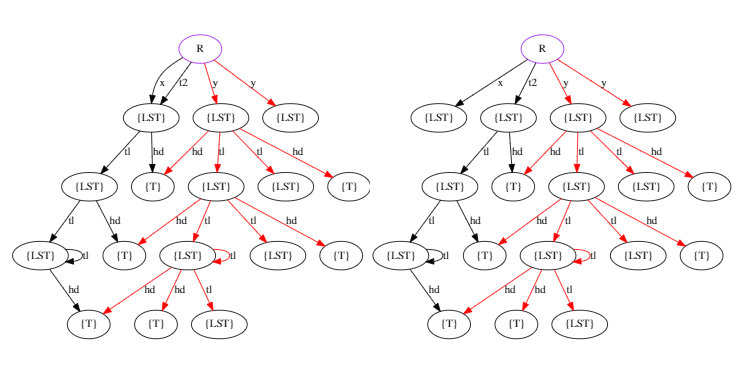}
\caption{(left) the Alias graph after executing program point \texttt{$L_2$} of Figure \ref{alg:b}. (right) the Alias graph after executing program point \texttt{$L_3$} of Figure \ref{alg:b}}
\label{fig:alias:graph}
\end{figure}

AutoAlias is at least as precise as the other approaches. All of the five properties are met by our implementation. What it is interesting is that other approaches fail to capture the fact that after the execution of routine \eif{Copy}, heads of $X$ and $Y$ might not be aliased at all. This is the case when the argument passed to \eif{Copy} is \eifkw{null}. According to \cite{Deutsch:1994}'s approach, the set of aliases of the algorithm in figure \ref{alg:a} is $\{(X\rightarrow (tl\rightarrow)^ihd, Y\rightarrow (tl\rightarrow)^jhd)\mid i=j\}$. For $i=0$ the set is $\{(X\rightarrow hd, Y\rightarrow hd)\}$, ruling out the possibility of no aliasing. AutoAlias captures this fact thanks to rule \texttt{AC-Cond} that analyses each branch of the conditional and unions the resulting alias diagrams, one of these diagrams yields no aliasing. Figure \ref{fig:alias:graph} depicts the respective alias graphs at different program points of the algorithm being analysed.

\section{Future work and conclusion}
\label{section:concl}
A widely available, widely applicable, easy-to-integrate and fast tool for alias analysis would immediately and immensely benefit many tasks of programming language implementation and verification. AutoAlias does not yet fulfill all these criteria but provides, in our opinion, a significant step forward. The examples to which we have applied to the tool so far, while still limited, provide encouraging evidence of the solidity and scalability of the approach. The application to change analysis, described in the companion paper, are currently the showcase, but many others are potentially open, of interest to both tool developers (in particular developers of compilers and verification tools) and application programmers.

We realize the extent of the work that remains ahead, including the following: taking into account tricky language mechanisms such as exceptions and function objects (closures in Java, delegates in C\#, agents in Eiffel); taking into account calls to external software mechanisms, e.g. system calls, which can potentially put the soundness of alias analysis into question since objects then go into the big bad world out there where anything can happen to them (but can we still reason about them without having to adopt the worst-case disaster scenario in which nothing can be assumed any longer?); refining the analysis and improving its precision further by taking into account ever more sophisticated patterns in conditional instructions and loops. 

It is our hope, however, that the present state of the work, as described in this article, advances the search for general and effective techniques of automatic alias analysis.

\section*{Acknowledgements}
We are indebted to colleagues who collaborated on the previous iterations of the Alias Calculus work, particular Sergey Velder (ITMO University) for many important suggestions regarding the theory, Alexander Kogtenkov (Eiffel Software, also then ETH Zurich) who implemented an earlier version of the Change Calculus, and Marco Trudel (then ETH Zurich). We thank members of the Software Engineering Laboratory at Innopolis University, particularly Manuel Mazzara and Alexander Naumchev, for many fruitful discussions.

\bibliographystyle{plain}
\bibliography{bibl}
\end{document}